\documentclass[aps,twocolumn,showpacs]{revtex4}
\usepackage{amsfonts}
\usepackage{amssymb,amsmath}
\pdfoutput=1
\usepackage{graphicx}
\def\he{\hbar/2e}

\begin{document}
%\begin{titlepage}
\title{Negative Mobility
 induced by  Colored  Thermal  Fluctuations}

\author{M. Kostur$^{1}$}
%\author{}
%\affiliation{Instituted  of Physics, University of Silesia, 40-007 Katowice, Poland }
\author{J. {\L}uczka$^{1}$}
%\author{}
%\affiliation{ Institute of Physics, University of Silesia, 40-007
%Katowice, Poland }
%\author{P. Talkner}
%\address{Institute of Physics, University of Augsburg, D-86135 Augsburg, Germany}
\author{P. H{\"a}nggi$^{2,3}$}
%\affiliation{Institute of Physics, University of Augsburg,
%D-86135 Augsburg, Germany}
%\author{Peter H\"anggi$^{1,3}$}
%\email{hanggi@physik.uni-augsburg.de}
\address{$1$  Institute of Physics, University of Silesia, 40-007
Katowice, Poland\\
$2$ Institut f\"ur Physik, Universit\"at Augsburg
Universit\"atsstr. 1, D-86135 Augsburg, Germany\\
 $3$ Department of Physics and Centre for Computational Science
 and Engineering, National University of Singapore, Republic of Singapore
117542}

\begin{abstract}
Anomalous transport of  non-Markovian, thermal Brownian particle
dynamics in spatially-periodic symmetric systems that is driven by
time-periodic symmetric driving and constant bias is investigated
numerically. The Brownian dynamics is modeled by a Generalized
Langevin equation with exponentially correlated Gaussian thermal
noise, obeying the fluctuation-dissipation theorem. We study the
role of non-zero correlation time of thermal fluctuations for the
occurrence of absolute negative (linear) mobility (ANM) near zero
bias, negative-valued, nonlinear mobility (NNM) and negative
differential mobility (NDM) at finite bias away from equilibrium. We
detect that a non-zero thermal correlation time can either enhance
or also diminish the value of ANM. Moreover, finite thermal noise
correlation can induce NDM and NNM in regions of parameter space for
which such ANM- and NNM-behavior is distinctly absent for limiting
white thermal noise. In parts of the parameter space, we find  a
complex structure of regions of linear and nonlinear negative
mobility: islands and tongues which emerge and vanish under
parameters manipulation. While certain such anomalous transport
regimes fade away with increasing temperature some specific regions
interestingly remain rather robust. Outside those regimes with
anomalous mobility, the ac/dc driven transport is either normal or
the driven Brownian particles are not transported at all.

\end{abstract}

%}
\maketitle
%
%%%%%%%%%%%%%%%%%%%%%%%%%%%%%%%%%%%%%%%%%%%%%%%%%%%%%%%%%%%%%%%%%%%%%%5

\section{Introduction}

Out-of-equilibrium transport processes are of prominent interest in
modern statistical physics from all, fundamental, experimental and
device  aspects \cite{hm,astumian}. A typical far from equilibrium
problem is generally nonlinear response of a system to an external
stimulus: For example when an external constant force $F_0$ acts on
the system and all other forces are of zero-average, it is expected
that the long-time, stationary average particle velocity $v$ becomes
an increasing function of the load $F_0$. An everyday example is
that of  a force pushing objects on a  table. A linear Ohmic
resistor characteristic constitutes  another example: An
increase of voltage is accompanied by a linear  increase of current.
This normal response behavior is distinct from cases with anomalous
transport. Familiar examples are the emergence of   negative
differential mobility  or conductivity, or the nonlinear
response involving negative(-valued) nonlinear mobility  away
from  the linear response behavior around {\it zero} applied
voltage. Here our focus is on yet another anomalous transport
behavior, namely so called ''absolute negative mobility''.
 This latter anomalous transport behavior refers to a regime
where the resulting velocity or current assumes the opposite sign of
the applied force or voltage around the zero bias regime. While the
NDM is common  for an abundance of nonlinear systems, the
phenomenon of ANM has been experimentally detected predominantly
much less frequently. Some examples that come to mind are the
nonlinear response in p-modulation-doped GaAs quantum wells
\cite{hop}, or also semiconductor superlattices \cite{keay}, and
recently also in driven Josephson junctions \cite{nagel}. The
phenomenon of ANM can typically relate to a genuine quantum effect,
involving asymmetry tunneling dynamics. In contrast, ANM as a result
of classical stochastic dynamics occurs more rare; but it is
expected to occur whenever stylized, ratchet-like structures
including geometric entropic barriers are present. A flurry of
recent theoretical \cite{eichhornPRL,eichhornPRE} and less few
experimental works \cite{nagel,ros} indeed prove such behavior.

The effect of ANM can occur also in the form of a far from
equilibrium phenomenon in driven nonlinear systems such as  in
nonlinear underdamped Brownian motion dynamics
\cite{prl_jj,prb_jj,reim} or even in overdamped nonlinear Brownian
motion in presence of time-delayed feedback \cite{hennig}.

In this work we shall focus on the case of  time-dependent driven
underdamped Brownian motion occurring in a periodic, reflection
symmetric potential and driven by thermal correlated noise. In prior
works \cite{prl_jj,prb_jj}, we have studied the  transport
properties of a classical Brownian particle of mass $m$ moving  in a
spatially periodic potential $V(x)=V(x+L)$ of period $L$ and barrier
height $\Delta V$, which is subjected to an external unbiased
time-periodic force $F(t)=F(t+T)$ of period $T= 2\pi / \Omega$ with
angular frequency $\Omega$ and of amplitude $A$. Additionally, a
constant bias $F_0$ acts on the system. This so defined Brownian
particle dynamics is then modeled by
%Newton's equation which is complemented by a noise term, i.e.
the   driven Langevin equation \cite{prl_jj}; i.e.,
\begin{eqnarray} \label{lan}
m \ddot x + \gamma \dot x = &-&V'(x) \nonumber\\ &+& A \cos(\Omega t + \phi_0)
 +F_0 +  \; \xi(t),
\end{eqnarray}
where $x=x(t)$ is a position of the particle at time $t$, a dot denotes differentiation
with respect to time and a
prime denotes a differentiation with respect to  the Brownian particle coordinate $x$.
 The parameter $\gamma$ denotes the viscous friction strength and  $\phi_0$ is an initial phase
 of the time-periodic driving. 
Here, the thermal fluctuations are modeled by  zero-mean, Gaussian
white noise $\xi(t)$ with the Dirac delta auto-correlation function
$\langle \xi(t)\xi(s)\rangle = 2 k_B T_0 \delta(t-s)$, where $k_B$ the Boltzmann constant and  $T_0$ denotes the temperature.

We could show that in the above system, there are distinct regimes
of anomalous transport. In particular  we can identify   both (i)
ANM and  (ii)  NDM and  the phenomenon of (iii)   negative nonlinear
mobility (NNM), occurring away from the   linear response regime
with respect to external bias $F_0$. We also remind the reader that
(\ref{lan}) mimics the physical realization for the behavior of a
physical  Josephson junction \cite{prl_jj,prb_jj,reim}.  In this
latter case, the periodic potential $V(x)$   has the  explicit
sinusoidal form, i.e.,
\begin{eqnarray}
\label{pot}
V(x) =  \Delta V \sin (2\pi x/L).
\end{eqnarray}
Notably, the theoretical findings for   ANM in  Josephson junctions
has recently  been verified  experimentally  with the work in Ref.
\cite{nagel}.

%%%%%%%%%%%%%%%%%%%%%%%%%%%%%%%%%%%%%%%%%%%%%%%%%%%%%%%%%%%%%%%%%%%%%%

%%%%%%%%%%%%%%%%%%%%%%%%%%%%%%%%%%%%%%%%%%%%%%%%%%%%%%%%%%%%%%%%%%%%%%%%%%%

\section{Periodically driven and biased non-Markovian Brownian Dynamics}

The thermal noise  $\xi(t)$ in Eq. (\ref{lan}) is approximated to
be ideally white noise with zero noise correlation time  $\tau_c =
0$. In real systems, however, the correlation time of thermal
fluctuations is only approximately zero. This approximation is
justified if $\tau_c$  is  much smaller than the smallest
characteristic time $\tau_s$ of the system itself. There are many
examples  where this situation is well satisfied  in real systems.
 There are, however, also situations where the thermal correlation time $\tau_c$
is of order or greater than $\tau_s$, so that the white-noise
approximation fails \cite{jungACP,chaos,goychuk2005}. In this latter
case a modeling based on the Markovian Langevin equation (\ref{lan})
is not correct;  instead, the generalized Langevin equation
should  then be invoked.

\subsection{The Generalized Langevin dynamics}

When the thermal noise is correlated, the appropriate Langevin
dynamics is a non-Markovian  dynamics  with
memory-friction described by  the so called Generalized Langevin Equation (GLE)
\cite{kubo66,zwan,JSP,LNP,RevMod}. It  explicitly reads
\begin{eqnarray}
\label{Gen}
m\ddot x(t) + \int_0^t K(t-s) \dot x(s)\;ds  = - U'(x(t), t) + \xi(t),
\end{eqnarray}
where the full potential  takes the form 
\begin{eqnarray}
\label{fullpot}
U(x, t) = V(x) - [A \cos(\Omega t +\phi_0)  +F_0] x.
\end{eqnarray}
This non-Markovian dynamics can be derived from first principles by
means of coupling the system of interest to a bath of harmonic
oscillators \cite{zwan,LNP} with the total system being prepared in
canonical thermal equilibrium \cite{LNP}. It then follows from the
central limit theorem that the  thermal fluctuations $\xi(t)$ obey a
zero-mean,  stationary typically non-Markovian Gaussian stochastic
process. The auto-correlation function of the thermal noise $\xi(t)$
is related to the memory (frictional) kernel $K(t)$ via the
fluctuation-dissipation relation \cite{kubo66,zwan,JSP,LNP,RevMod}:
\begin{eqnarray}
\label{mom}
\langle \xi(t) \xi(s)\rangle = k_BT_0 K(|t-s|).
\end{eqnarray}
 Interestingly, due to the nonlinearity of the potential $V(x)$ it
 is still an unsolved,  open problem to derive the explicit form of the
 generalized  master equation for the single-event non-Markovian
 probability $p(x, \dot x,t)$, see in Refs. \cite{JSP,LNP,ZPhysik};
 this task is achieved only in form of a time-convolutionless master equation
 with time-dependent transport coefficients
 iff the potential $V(x)$ is at most quadratic in $x$ only
 \cite{LNP,ZPhysik}; then yielding  a general non-Markovian Gauss
 process for the equilibrium dynamics $(x(t),\dot x (t))$.
%%%%%%%%%%%%%%%%%%%%%%%%%%%%%%%%%%%%%%%%%%%%%%%%%%%%%%%%%%%%%%%%%%%%%%%%%%%%

%%%%%%%%%%%%%%%%%%%%%%%%%%%%%%%%%%%%%%%%%%%%%%%%%%%%%%%%%%%%%%%%%%%%%%%%%

\subsection{Exponentially correlated  thermal fluctuations}

The Gaussian thermal fluctuations $\xi(t)$ in (\ref{Gen})  are
completely determined by the memory function $K(t)$.  If the memory
function $K(t)$  is the Dirac delta function, i.e.,  $K(t)=2\gamma
\delta(t)$, then Eq. (\ref{Gen}) reduces to the form (\ref{lan}). A
well studied form of   correlated fluctuations   is defined by means
of an Ornstein-Uhlenbeck (O-U) stationary stochastic process for
$\xi(t)$ \cite{chaos}.  This Gaussian Markov process for the thermal
noise (note that the resulting  Brownian dynamics is then still
non-Markovian) is henceforth  correlated exponentially. We next
use a Markovian embedding of the GLE dynamics in Eq. (\ref{Gen}).
Towards this objective we   present the correlation function in
the form
\begin{eqnarray}  \label{o-u}
\langle \xi(t) \xi(s)\rangle = k_BT_0 K(|t-s|)
=\frac{\gamma k_BT_0}{\tau_c} \;\mbox{e}^{-|t-s|/\tau_c},
\end{eqnarray}
where $\tau_c$ is the correlation time of the  O-U process 
 which we can then smoothly vary from the limit of white
Gaussian noise ($\tau_c =0$) to strongly correlated thermal noise
($\tau_c \gg \tau_0$).
Because the integral kernel exhibits an exponential form,  we can
convert Eq. (\ref{Gen}) into a set of ordinary stochastic
differential equations: Let us define  the auxiliary stochastic
process $w(t)$ via the relation
\begin{eqnarray}
\label{y(t)}
w(t) = \frac{\gamma}{\tau_c} \int_0^t \mbox{e}^{-(t-s)/\tau_c}\dot x(s)\;ds.
\end{eqnarray}
%
%which is the integral of Eq. (\ref{Gen}).
Then Eq. (\ref{Gen}) is equivalently  transformed  into the form
\begin{eqnarray}
\label{4eq}
m \dot v(t)&=& - U'(x(t), t) -w(t) + \xi (t),   \\
\dot x(t)&=&v(t),  \\
\dot w(t)&=& -\frac{1}{\tau_c} w(t) +\frac{\gamma}{\tau_c} v(t),  \\
\label{O-U}
\dot \xi (t)&=& -\frac{1}{\tau_c} \xi (t)
+\frac{1}{\tau_c} \sqrt{2\gamma k_BT_0} \; \Gamma(t),
\end{eqnarray}
where the normalized white  Gaussian noise  $\Gamma(t)$ obeys
 $\langle\Gamma(t)\Gamma(s)\rangle=\delta(t-s)$ while the
last equation of this set describes the O-U noise with the
exponential correlation function (\ref{o-u}) \cite{pha,jungprl}.
Note  that in Eq. (\ref{4eq}), the linear combination $z(t) = \xi
(t)-w(t)$ occurs. By subtracting the two last relation we then find
the set of three coupled Markovian Langevin equations, modeling the
two-dimensional  non-Markovian GLE in Eq. (\ref{Gen}); i.e.,
\begin{eqnarray}
\dot x(t) &=& v(t),
\label{x}\\
\label{v}
 \dot v(t) &=&  -\frac{1}{m} U'(x(t), t) +\frac{1}{m} z(t), \\
\dot z(t) &=& -\frac{1}{\tau_c}z(t) - \frac{\gamma}{\tau_c} v(t)
+\frac{1}{\tau_c} \sqrt{2\gamma k_BT_0}\; \Gamma(t).
\label{z}
\end{eqnarray}
The corresponding three-dimensional Fokker-Planck equation for $p(x,
v, z, t)$ is numerically cumbersome to implement for $V(x$) a
nonlinear, spatially periodic function. Alternatively we apply
direct numerical methods for the solution of the three coupled
Langevin equations in Eqs (\ref{x})-(\ref{z}).

%%%%%%%%%%%%%%%%%%%%%%%%%%%%%%%%%%%%%%%%%%%%%%%%%%%%%%

The  limiting case of white noise has been analyzed in detail in
Refs \cite{prl_jj,prb_jj}, where a numerical method  has been
described, see also  in Refs \cite{JPC1,JPC2}. In the remaining of this
work we shall use the dimensionless form of  the set of equations
(\ref{x})-(\ref{z}). In doing so we  scale  coordinate $x$ and time
$t$ as follows
\begin{eqnarray}
\label{scaling}
X= \frac{x}{L}, \qquad \hat{t} = \frac{t} {\tau_0},
 \qquad \tau_0^2 = \frac{mL^2}{\Delta V}.
\end{eqnarray}
Then, the set of equations (\ref{x})-(\ref{z}) is recast as
\begin{eqnarray}
\label{X}
\dot X &=&  Y    \\
\label{Y}
\dot Y &=&  - W'(X) + a \cos(\omega \hat{t}+ \phi_0) + f + Z,    \\
\dot Z &=& -\frac{1}{\hat{\tau_c}} Z - \frac{\hat{\gamma}}{\hat{\tau_c}} Y
+\frac{1}{\hat{\tau_c}} \sqrt{2\hat{\gamma} D}\;  \hat{\xi}(\hat{t}),
\label{Z}
\end{eqnarray}
where
\begin{eqnarray}
\label{YZ}
Y = \frac{\tau_0 }{L}\; v, \qquad Z  = \frac{L} {\Delta V} \; z
\end{eqnarray}
and a dot denotes differentiation with respect to the re-scaled time
$\hat{t}$. Here, $Y=Y({\hat t})$  is the  dimensionless velocity of
the Brownian particle and  $Z=Z({\hat t})$ denotes the corresponding
dimensionless random force. The remaining re-scaled parameters are:
(1) the friction coefficient $\hat{\gamma} = (\gamma / m) \tau_0 =
\tau_0 / \tau_L$ equals the ratio of  two characteristic times,
namely time $\tau_0$ and the relaxation time  of the velocity degree
of freedom, i.e. $\tau_L = m/\gamma$; (2) the potential
$W(X)=V(x)/\Delta V = W(X+1) = \sin (2\pi X)$ possesses unit period
and barrier height $\Delta \hat{V}=2$; (3) the amplitude $a = L A /
\Delta V$ and the frequency $\omega = \Omega \tau_0$ (or the period
${\cal T}=2\pi/\omega$); (4) the load $f=L F_0/ \Delta V$; (5)
 the zero-mean white noise
$\hat{\xi}(\hat{t})$ is correlated as
$\langle\hat{\xi}(\hat{t})\hat{\xi}(\hat{s})\rangle=\delta(\hat{t}-\hat{s})$
with a re-scaled noise intensity $D = k_B T_0 / \Delta V$.  The
latter is   given as the ratio of two energies, the thermal energy
and the half of barrier height of the  potential $V(x)$.

From here on,  we shall  use only these dimensionless variables and
shall omit the  notation``hat'' in all quantities in Eqs
(\ref{X})-(\ref{Z}).

%%%%%%%%%%%%%%%%%%%%%%%%%%%%%%%%%%%%%%%%%%%%%%%%%%

%%%%%%%%%%%%%%%%%%%%%%%%%%%%%%%
\begin{figure}[htpb]
  \begin{center}
\includegraphics[width=0.45\textwidth,angle=0]{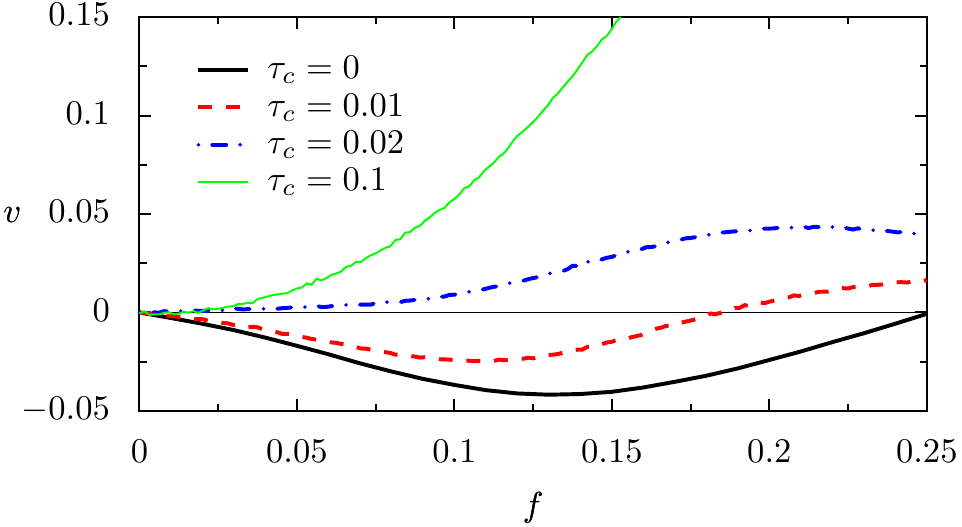}
\end{center}
  \caption{(color online) Regime of  destructive role of finite thermal
  noise correlations  on absolute negative mobility (ANM):
The averaged long-time velocity $v$ of the ac  driven Brownian
particle is depicted as a
  function of the  externally applied, static force $f$. The chosen system parameters
  are: $a=4.578$, $\omega=4.9$, $\gamma=0.9$ and $D=0.001$.  While
  white thermal noise ($\tau_c=0$) does depict ANM this is increasingly
  diminished upon increasing the  correlation time of thermal noise
  until  it is undone into a normal, positive mobility behavior.
}
  \label{fig1}
\end{figure}
%%%%%%%%%%%%%%%%%%%%%%%%%

\section{Numerical Results for Brownian non-Markovian  Anomalous Transport}

We next study numerically the long-time transport characteristics of
the nonlinear Brownian dynamics with memory friction. In particular
we shall focus on the current feature as given by  the  long-time
averaged velocity $v\equiv \langle Y \rangle$. This averaging is
performed as follows: First we  perform an average over all
realizations of the thermal fluctuations which yields a temporally
varying quantity. A second temporal average is the cycle period of
the external ac-driving.  Because the
resulting asymptotic, long time dynamics is not necessarily ergodic
(i.e. independent of chosen initial conditions) in all phase space
we also need to perform an average over unbiased initial conditions.
We have chosen uniformly distributed initial positions $X(t=0)=X_0$
over one period of the periodic potential $W(X)$; the initial
velocities $Y(t=0)=Y_0$ are unbiased and taken as uniformly
distributed in the interval $[-2, 2]$ and the initial phase $\phi_0
\in [0, 2\pi]$. The symmetry consideration of Eq. (\ref{Gen}) then
implies that this average velocity $v(f)$ as a function of the
external constant force $f$ is an odd function, i.e., $v(-f)=-v(f)$;
thus $v(f=0)=0$. Therefore we will in our numerics only consider the
half-axis with $f \ge 0$.

Because, as shown with previous works with white thermal noise
\cite{prl_jj,prb_jj,reim}, the transport dynamics becomes very rich
indeed in all parameter space exhibiting all, namely ANM, NDM and
NNM. As it must be expected this richness does not diminish with yet
another parameter of variation, namely  the correlation time of
thermal noise $\tau_c$. Also it must be kept in mind that it is
impossible to scan numerically over {\it all} possible parameter
space of the driven nonlinear Brownian dynamics. In the following we
shall focus our numerical study to regimes in parameter space that
are (i) experimentally accessible \cite{prl_jj,prb_jj} and (ii)
exhibit a most interesting complexity for the nonlinear response.

Clearly, the resulting velocity $v=v(f)$ is typically nonlinear in
external bias $f$. The linear response behavior is defined for small
bias $f\rightarrow 0$ as
\begin{equation}
\label{mob}
v(f\rightarrow 0) = \mu f,
\end{equation}
where the (linear)  mobility $\mu$ can become negative, $\mu < 0$.
This regime will be termed    absolute negative mobility (ANM); i.e.
an anomalous transport regime for which the the particle is
transported in the {\it opposite direction} to the externally
applied force $f$. Moreover, negative nonlinear mobility (NNM)
refers to an anomalous transport regime for which we find  $v(f)/f
<0$ in some finite intervals of $f$, being disjoint from the
interval around $f=0$ . Finally, regimes  of  negative differential
mobility (NDM), when $dv(f)/df < 0$ in some intervals of $f$, can be
detected.

\subsection{Controlling ANM with noise correlation time}
\subsubsection{Undoing ANM with small noise correlation time}

We first consider small bias values $f$ and study the linear
($f\rightarrow 0$) response and the accompanying nonlinear response
with increasing $f$ to larger values for specific parameter
settings. We first zoom into a parameter regime for which we find
ANM for $\tau_c=0$. An example is depicted with Fig. 1. If the
correlation time $\tau_c$ increases, starting out from zero, we
 observe a diminishing ANM with increasing $\tau_c$, until it disappears
 and turns into normal, positive-valued mobility upon increasing $\tau_c$ further.  Put differently, the
mobility coefficient $\mu$ in Eq. (\ref{mob}) starts to increase
from negative values, passes through zero and eventually becomes
positive. For  $\tau_c > 0.035$, the velocity monotonically
increases in the region of  small bias $f$ with $\mu > 0$. Note that
the dimensionless correlation time is rather short in comparison
 to the other   characteristic time scales, that is  with the characteristic  time-scale $\tau_0= 1$,
the characteristic velocity relaxation time-scale is $1.1 \tau_0$
and time of periodic driving is  ${\cal T} =1.28 \tau_0$\/.

A first main finding therefore is that even a small, finite thermal
noise correlation  time can diminish and even undo ANM.

%%%%%%%%%%%%%%%%%%%%%%%%%%%%%%%
\begin{figure}[htpb]
  \begin{center}
\includegraphics[width=0.42\textwidth,angle=0]{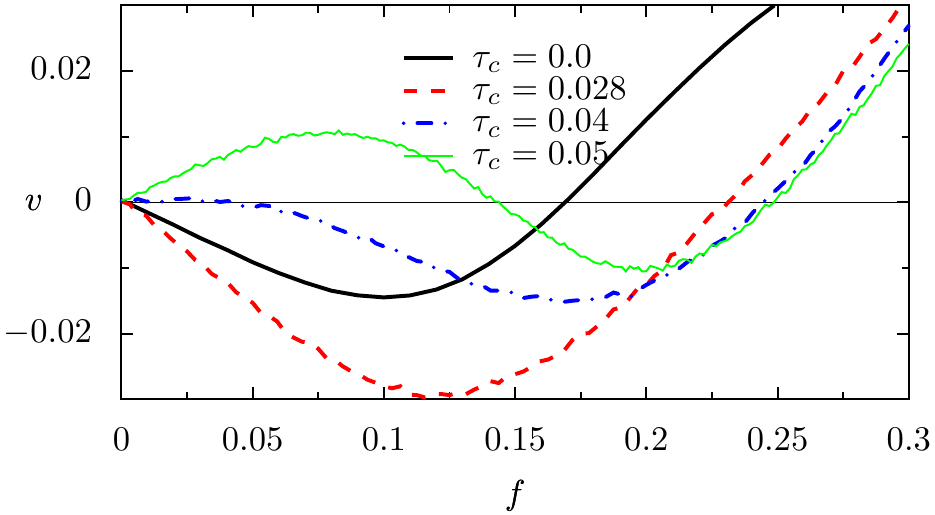}
\includegraphics[width=0.42\textwidth,angle=0]{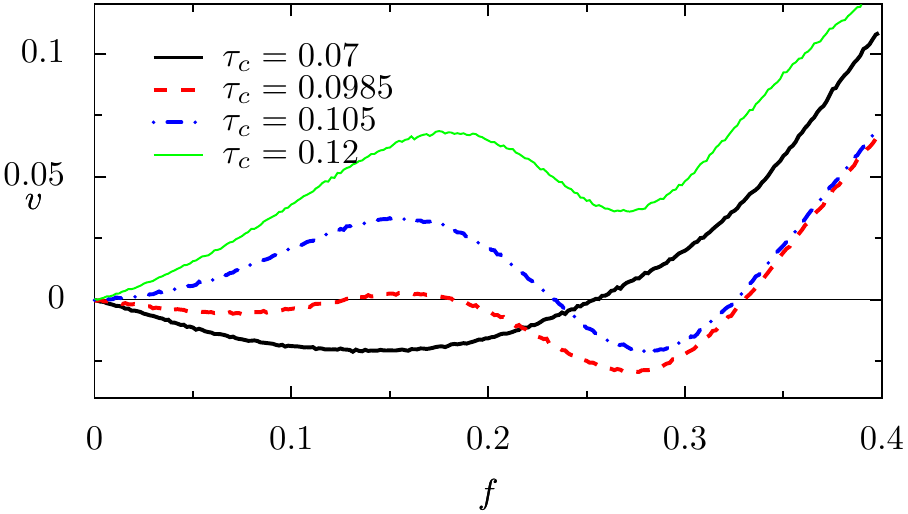}
\end{center}
  \caption{(color online)  Regime of   constructive role of thermal noise correlations
on absolute negative mobility: The averaged long-time velocity $v$
of the Brownian particle is depicted as a
  function of the  bias force $f$ for selected values of the noise correlation time $\tau_c$  of thermal fluctuations.
The  system parameters are chosen  different from Fig. \ref{fig1}
and read: $a=4.1293$, $\omega=4.9$, $\gamma=0.9$ and $D=0.001$.
Panel (a):  Within this regime we find that  a small thermal noise
correlation time can {\it enhance} ANM.  Panel (b): Upon further
increasing thermal noise correlation time $\tau_c$ from the values
 shown in (a) we observe a turnover into ANM again, note the case  with $\tau_c
=0.07$; followed by a NNM behavior, see $\tau_c=0.0985$,   and
finally a re-entrance into a normal transport regime is observed at
small bias with a regime following  with NDM as $f$ is increased
further; note the case with $\tau_c=0.12$. }
\label{fig2}
\end{figure}
%%%%%%%%%%%%%%%%%%%%%%%%%%%%%%%%%%%%%%%%%%%%%%%%

\subsubsection{Enhancing ANM and creating NNM}

Upon scanning the parameter set to a different forcing strength of
the ac-driving we find that thermal noise correlations can in fact
also {\it enhance} rather than diminish the value of ANM. This is
illustrated in Fig. \ref{fig2}, panel (a), for noise correlation
times varying in the interval $\tau_c\in [0, 0.028]$. Beyond
$\tau_c=0.028$  ANM diminishes again, exhibiting as well regimes
with NNM,  cf. the case $\tau_c=0.05$.

It is intriguing to note that within this parameter setting  a
further increase of the correlation time yields an opposite
behavior, see panel (b) in Fig. 2. The mobility coefficient $\mu$
then is   decreasing  from positive values and next turns into ANM
again.  The most pronounced ANM-value occurs around around
$\tau_c=0.08$ (not shown) before entering a regime with coexistence
with both  ANM and NNM around $\tau_c=0.1$.  For larger correlation
time we find normal linear mobility followed up in the nonlinear
regime with a region exhibiting NDM.

% too much detail / complexity for a "normal" reader --he will guess this
%Yet other scenarios   can  also be identified: For example, first
%the NNM start to shrink and finally ANM vanishes.

We  note that the various situations  described above do occur and
can coexist in other settings of the parameters.

%%%%%%%%%%%%%%%%%%%%%%%%%%%%%%%
\begin{figure}[htpb]
  \begin{center}
\includegraphics[width=0.42\textwidth,angle=0]{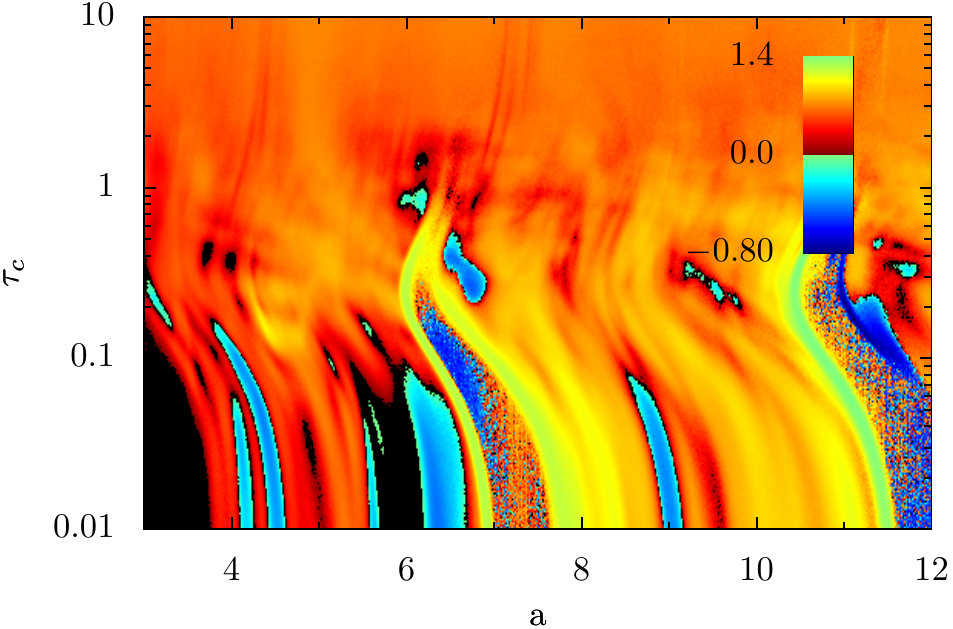}
\includegraphics[width=0.45\textwidth,angle=0]{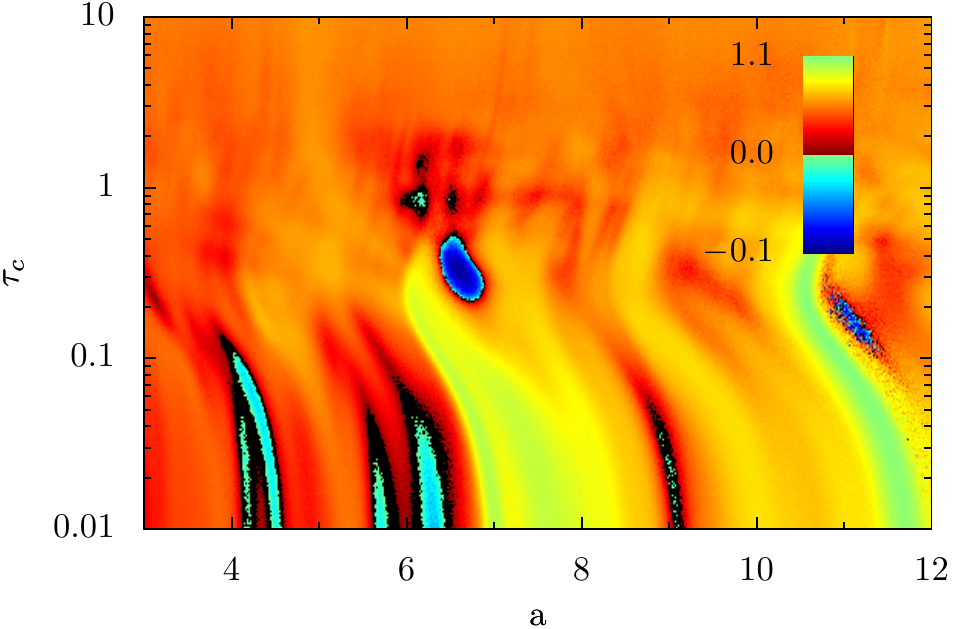}
\includegraphics[width=0.45\textwidth,angle=0]{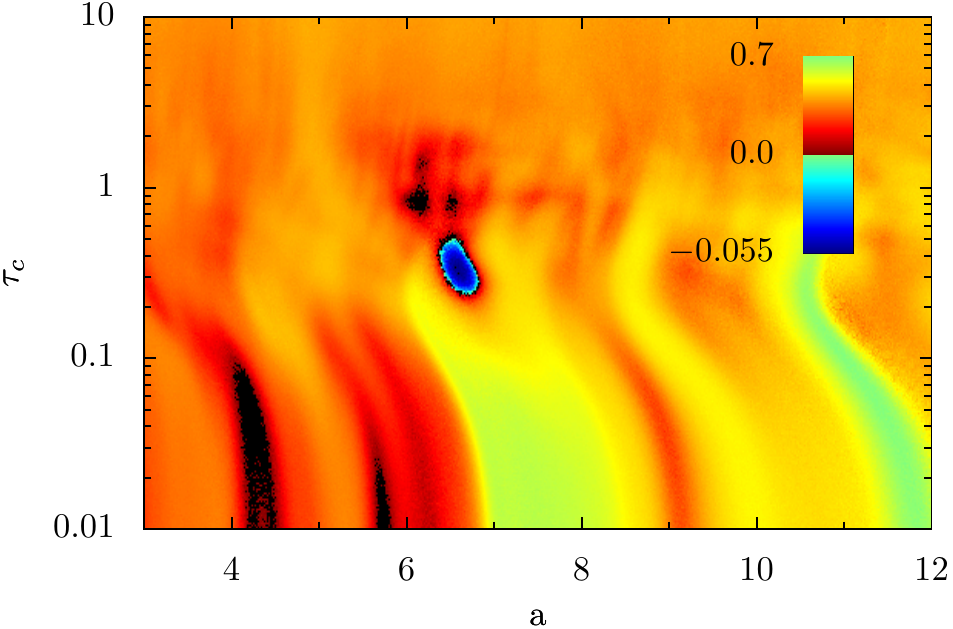}
\end{center}
  \caption{(color online)
Regions of negative (nonlinear) mobility manifested either as ANM or
also NNM, in the parameter plane given by $\{a, \tau_c\}$ for three
different temperatures, $D=0.0001, 0.001, 0.01$ (top to bottom). The
remaining parameters are fixed: namely $a=4.1293$, $\omega=4.9$,
$\gamma=0.9$ and $f=0.1$. }
  \label{fig3}
\end{figure}
%%%%%%%%%%%%%%%%%%%%%%%%%

 \subsection{Correlation time induced Islands and Tongues of ANM and NNM}

While in the previous subsection we have varied the external bias
$f$ away from zero we next keep this bias {\it fixed} at $f=0.1$ and
scan instead the correlation time $\tau_c$ versus the ac-driving
strength $a$. The emerging asymptotic, averaged velocities are then
depicted in color-coded plots as shown with Fig. \ref{fig3} for
three different settings of the temperature $T_0$.

Not unexpected, there occur in this parameter space several domains
with  anomalous transport features. The  geometric structure of
these domains in the depicted  regime of the $\{a, \tau_c\}$-variation is very complex. Let us remind ourselves that
the underlying deterministic dynamics is chaotic in some regimes and
therefore a fractal structures of certain domains must be expected
to exist. We are interested in the stability of those domains in
parameter space for which  ANM and/or NNM occur.

Our comprehensive numerical investigations reveal a rich diversity
of structures, formed by regions of both, ANM  and NNM. We present
one example and visualize it in the parameter-plane $\{a, \tau_c\}$.
In Fig. 3, we depict how the plane is divided into regions of normal
($v(f) > 0$ for $f>0$ ) and anomalous ($v(f) < 0$  for $f>0$)
transport.  In doing so we here do not discriminate between ANM and
NNM. Both these transport behaviors are jointly presented. Two
thermal situations, namely, 'low' and 'high' temperatures are shown
for comparison. At 'low' temperature, cf. panel (a),  we reveal the
refined structure with many narrow, slim and twisted regions of
ANM-NNM. Some of those regions, the 'tongues', survive with
correlation time $\tau_c$ approaching zero  (i.e. the horizontal
abscissa) and there are 'islands' of negative mobility which
disappear for $\tau_c=0$. If one  fixes one of the parameters, say
$\tau_c=0.1$, intervals of negative mobility are clearly noticeable:
there are several intervals of the amplitude $a \in (a_i, a_{i+1})$
for which ANM-NNM can be  detected. Outside these intervals, a
normal response to the load $f$ is found.

If temperature is increased, this small temperature structure is
increasingly washed out; i.e. it  becomes more smooth. Many
previously existing domains of ANM-NNM behavior start to shrink or
vanish altogether. We detect some few robust regimes for which
anomalous transport persists, namely a few islands and a few
tongues. The most robust such island against increasing noise
strength (temperature) is the island located around  $\{a, \tau_c\}
= \{6.56, 0.31\}$.  This very domain even survives at 'high'
temperatures.

We emphasize that such complicated regimes of ANM-NNM are not just
rare occurrences: they can be verified with numerically
arbitrarily-high-accuracy  calculations and over extended intervals
in parameter space. Given the complexity of the underlying dynamics
with time-dependent ac-driving, nonlinearity and in presence of
noise with finite correlation time the observed behavior is clearly
beyond a sensible  analytical description.

\subsection{Current-voltage characteristics of a realistic Josephson junction device}

As an application of the above theoretical study we consider next a
Josephson junction for which the anomalous conductance have been
measured  in Ref. \cite{nagel}. The relation in Eq. (\ref{lan}) with
the potential (\ref{pot})  models the resistively and capacitively
shunted Josephson junction, also known as  the Stewart-McCumber
model \cite{stewart,mccumber,kautz}. It contains three additive
current contributions: a Cooper pair tunnel current characterized by
the critical current $I_0$, a normal (Ohmic) current characterized
by the normal state resistance $R$ and a displacement current due to
the capacitance $C$ of the junction. For this model,  the position
$x$ of the Brownian particle translates into  is the phase
difference $\phi$  between the macroscopic wave functions of the
Cooper pairs on both sides of the junction, i.e., $x = \phi$, the
mass $m=(\he)^2 C$,   the friction coefficient $\gamma =
(\he)^2(1/R)$, the barrier hight $\Delta V = (\he) I_0$ and the
period $L=2\pi$. The load $F_0=(\he) I_{d}$ is given by means of the
dc-bias current, the amplitude $A=(\he) I_{a}$ and the frequency
$\Omega$ define the external ac current. The velocity $v=\dot x$
translates into the voltage across the junction.
%%%%%%%%%%%%%%%%%%%%%%%%%%%%%%%
\begin{figure}[htpb]
  \begin{center}
\includegraphics[width=0.45\textwidth,angle=0]{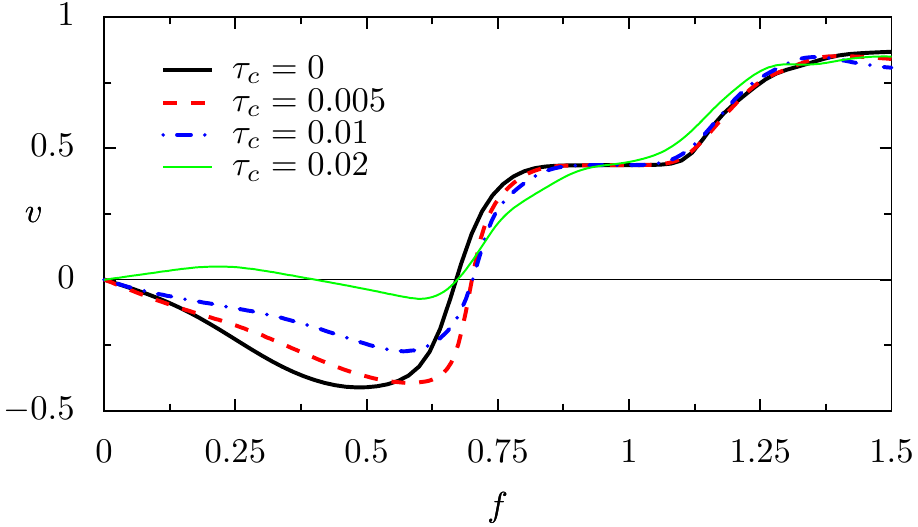}
\end{center}
  \caption{(color online)
Role of  non-zero  thermal
  noise correlations  on absolute negative mobility
 for a  parameter set corresponding to the anomalous Josephson junction transport
 regime used in the experiment in Ref. \cite{nagel}.
The  long-time averaged voltage  $v$ across the junction is depicted
as as a function of the  externally applied dc current $f$. The
corresponding dimensionless parameters are: $a=13.874$,
$\omega=2.75$, $\gamma=2.264$ and $D=0.000895$.
  The white thermal noise regime (i.e. $\tau_c=0$)  depicts ANM-behavior which  is
  sustained
  for  short  correlation times of the thermal noise.
For increasingly larger correlation times ANM is diminished and
eventually  vanishes all together and crosses over into  a normal,
 positive mobility behavior.  Still,  negative-valued, nonlinear mobility can occur  far away from
 equilibrium; note the nonlinear response behavior for $\tau= 0.02$ and around   $f \sim
0.6 $. } \label{fig1}
\end{figure}
%%%%%%%%%%%%%%%%%%%%%%%%%
The experimental results for the Josephson junction presented  in
Ref. \cite{nagel} show very good agreement with the Stewart-McCumber
model with the   following set of parameter values: namely, the
dimensionless amplitude of the ac-current $a= 13.874$, the frequency
$\omega =2.75$, the friction coefficient $\gamma =2.264$ and the
(white) noise intensity $D=0.000895$.
 We note that Eq. (\ref{Gen})  can serve as a generalization
of the Stewart-McCumber model towards a regime  of validity to lower
temperatures where finite correlations of the thermal fluctuations
increasingly play a significant role \cite{kautz}. We numerically
find that at small non-zero correlation time of the thermal
fluctuations ANM is sustained within  tailored parameter regimes, as
studied previously in the context of current-voltage characteristics
of a ac driven Josephson junction device in  Ref. \cite{nagel}, see
Fig. 4. With increasing thermal noise correlation time ANM is
weakened and finally turns into normal behavior followed by a regime
of NNM far away from equilibrium.  This feature of finite thermal
noise correlation is of relevance from an experimental point of view
when measuring mobility as a function of temperature.  We can also
conclude from this study that at the temperature of the experiment
performed in Ref. \cite{nagel},   the white thermal noise
approximation is seemingly well satisfied.

\section{Conclusions}

With this study we numerically  analyzed the role of non-zero
correlation time of thermal fluctuations on the anomalous transport
regimes of underdamped, non-Markovian Brownian particles that are
driven by time-periodic and static forces. We detected a rich
variety of anomalous transport behavior in an experimentally wide
parameter space where anomalous transport can be monitored. The
regions of absolute negative mobility and negative nonlinear
mobility form complicated structures in parameter space with
stripes, fibres and islands, see Fig. 3. At low temperatures these
structures  can unambiguously be attributed to finite correlations
of  thermal fluctuations. Such correlations are either constructive
or destructive in nature with respect to the size of anomalous
transport coefficients such as mobility. A subsequent increase of
temperature starts to blur these structure. Nevertheless, there
occur stable 'islands' with anomalous, negative-valued mobility
behavior which are solely induced by non-zero thermal noise
correlations, cf. Fig. 3.

In the regime of linear mobility response, such  non-zero thermal
correlations are found either to diminish or also to enhance the
value of absolute negative (linear) mobility, dependent on the
specific parameter setting for ac-driving strength and /or remaining
parameters.

We also have compared our predictions in a parameter regime of a
recent experiment on anomalous response behavior in a ac/dc driven
Josephson junction \cite{nagel}: The observed ANM behavior is
sustained  for small thermal noise correlation time but increasingly
fades out with increasing thermal noise correlation, see in Fig. 4.

\section*{Acknowledgments}
Work  supported by DAAD (J. {\L}.), the Polish Ministry of Science and Higher
Education under the grant  N 202 203 534,  the German Excellence
Initiative via the  ''Nanosystems Initiative Munich (NIM)'' and by
the DFG through the collaborative research center SFB-486.

 \end{document}